\documentclass[10pt, conference, compsocconf]{IEEEtran}

\usepackage{graphics}
\usepackage{epsfig}
\usepackage{epstopdf}

\usepackage{graphics}
\ifCLASSINFOpdf
\else
\fi
\begin{document}
%
% paper title
% can use linebreaks \\ within to get better formatting as desired
\title{Identifying Influential Bloggers: Time Does Matter}

% author names and affiliations
% use a multiple column layout for up to two different
% affiliations

\author{\IEEEauthorblockN{Leonidas Akritidis, Dimitrios Katsaros, Panayiotis Bozanis}
\IEEEauthorblockA{Department of Computer \& Communication Engineering\\
University of Thessaly\\
Volos, Greece\\
\{leoakr, dkatsar, pbozanis\}@inf.uth.gr}

}

% conference papers do not typically use \thanks and this command
% is locked out in conference mode. If really needed, such as for
% the acknowledgment of grants, issue a \IEEEoverridecommandlockouts
% after \documentclass

% for over three affiliations, or if they all won't fit within the width
% of the page, use this alternative format:
% 
%\author{\IEEEauthorblockN{Michael Shell\IEEEauthorrefmark{1},
%Homer Simpson\IEEEauthorrefmark{2},
%James Kirk\IEEEauthorrefmark{3}, 
%Montgomery Scott\IEEEauthorrefmark{3} and
%Eldon Tyrell\IEEEauthorrefmark{4}}
%\IEEEauthorblockA{\IEEEauthorrefmark{1}School of Electrical and Computer Engineering\\
%Georgia Institute of Technology,
%Atlanta, Georgia 30332--0250\\ Email: see http://www.michaelshell.org/contact.html}
%\IEEEauthorblockA{\IEEEauthorrefmark{2}Twentieth Century Fox, Springfield, USA\\
%Email: homer@thesimpsons.com}
%\IEEEauthorblockA{\IEEEauthorrefmark{3}Starfleet Academy, San Francisco, California 96678-2391\\
%Telephone: (800) 555--1212, Fax: (888) 555--1212}
%\IEEEauthorblockA{\IEEEauthorrefmark{4}Tyrell Inc., 123 Replicant Street, Los Angeles, California 90210--4321}}

% use for special paper notices
%\IEEEspecialpapernotice{(Invited Paper)}

% make the title area
\maketitle

\begin{abstract}
Blogs have recently become one of the most favored services on the Web. Many users maintain a blog and write posts to express their 
opinion, experience and knowledge about a product, an event and every subject of general or specific interest. More users visit blogs 
to read these posts and comment them. This ``participatory journalism'' of blogs has such an impact upon the masses that Keller and 
Berry argued that through blogging ``one American in tens tells the other nine how to vote, where to eat and what to buy''~\cite{keller1}. Therefore, 
a significant issue is how to identify such influential bloggers. This problem is very new and the relevant literature lacks 
sophisticated solutions, but most importantly these solutions have not taken into account temporal aspects for identifying influential 
bloggers, even though the time is the most critical aspect of the Blogosphere.
This article investigates the issue of identifying influential bloggers by proposing two easily computed blogger ranking methods, which incorporate 
temporal aspects of the blogging activity. Each method is based on a specific metric to score the blogger's posts. The first metric,
termed MEIBI, takes into consideration the number of the blog post's inlinks and its comments, along with the publication date of the
post. The second metric, MEIBIX, is used to score a blog post according to the number and age of the blog post's inlinks and its comments.
These methods are evaluated against the state-of-the-art influential blogger identification method utilizing data collected from a 
real-world community blog site. The obtained results attest that the new methods are able to better identify significant temporal patterns 
in the blogging behaviour.

\end{abstract}

\begin{IEEEkeywords}
Blogosphere; influential bloggers; ranking

\end{IEEEkeywords}

% For peer review papers, you can put extra information on the cover
% page as needed:
% \ifCLASSOPTIONpeerreview
% \begin{center} \bfseries EDICS Category: 3-BBND \end{center}
% \fi
%
% For peerreview papers, this IEEEtran command inserts a page break and
% creates the second title. It will be ignored for other modes.
\IEEEpeerreviewmaketitle

\section{Introduction}
% no \IEEEPARstart
During the last years, we have witnessed a massive transition in the applications and services hosted on the Web. The obsolete static Web
sites have been replaced by numerous novel, interactive services whose common feature is their dynamic content. The social and
participatory characteristics that were included in these services, led to the generation of virtual
communities, where users share their ideas, knowledge, experience, opinions and even media content. Examples include blogs, forums,
wikis, media sharing, bookmarks sharing and many others, which are collectively known as the Web 2.0.

%% In particular: We deal with the Blogosphere, and specifically with community blogs
Blogs are locations on the Web where individuals (the bloggers) express opinions or experiences about a subject. Such entries are called 
blog posts and may contain text, images, embedded videos or sounds and hyperlinks to other blog posts and Web pages. On the other hand,
the readers are provided with the ability to submit their own comments in order to express their agreement or disagreement to the ideas
or opinions contained in the blog post. The comments are usually placed below the post, displayed in reverse chronological order. The
virtual universe that contains all blogs is known as the \textit{Blogosphere} and accommodates two types of blogs~\cite{al08}: a) {\it
individual blogs}, maintained and updated by one blogger (the blog owner), and b) {\it community blogs}, or multi-authored blogs, where
several bloggers may start discussions about a product or event. Since in the former type of blogs, only the owner can start a new
line of posts, the present article focuses only on community blogs.

%% The specific problem we are interested in: The Influential bloggers
In a physical community, people use to consult others about a variety of issues such as which restaurant to 
choose, which medication to buy, which place to visit or which movie to watch. Similarly, the Blogosphere is a virtual world where
bloggers buy, travel and make decisions after they listen to the opinions, knowledge, suggestions and experience of other bloggers.
Hence, they are \textit{influenced} by others in their decision making and these others are defined in~\cite{keller1} as~\textit{the
influentials}.

%% Is the problem we investigate singifincant? Is the motivation of our paper significant?
The identification of the influentials is of significant importance, because they are usually connected in large virtual communities and 
thus they can play a special role in many ways. For instance, commercial companies can turn their interest in gaining the respect of the
influentials to become their ``unofficial spokesmen", instead of spending huge amounts of money and time to advertise their products
to thousands of other potential customers. It can also lead to the development of innovative business opportunities (related to commercial 
transactions and travelling), can assist in finding significant blog posts~\cite{eacc08,hmo08}, and can even be used to influence other 
peoples' voting behavior.

%% Briefly introduce our novelty, downvaluing relevant work
The issue of identifying influential bloggers is very recent and despite it seems similar to problems like the
identification of influential blog sites~\cite{gill1} and the identification of authoritative Web
pages~\cite{Langville-Google-book-2006}, the techniques proposed for these problems can not be applied to the identification of
influential bloggers. The problem of identifying the influential bloggers has been introduced in~\cite{agarwal2}, and the
literature lacks other sophisticated solutions. That initial model, mentioned here as {\it the influence flow method}, explicitly
discriminated the influential from the active (i.e., productive) bloggers, and considered features specific to the Blogosphere,
like the blog post's size, the number of comments, and the incoming and outgoing links. Nevertheless, this model fails to incorporate temporal aspects which are crucial to the Blogosphere and does not take into account the productivity as another factor which affects the influence.

%% The paper's contribution
Motivated by these observations, this article proposes a new way of identifying influential bloggers in community blogs, by considering both the temporal and productivity aspects of the blogging behavior, along with the inter-linkage among the blogs posts. The proposed 
methods are evaluated against the aforementioned initial model (which is the only competitor so far) using data from a real-world blog site.

%% The paper's roadmap
The rest of the paper is organized as follows: In Section~\ref{sec-related-work} we briefly present the relevant work, describing in more
details the only method which is closely relevant to the problem considered here. Section~\ref{sec-scientometrics} introduces the proposed
algorithms for the identification of influential bloggers; in Section~\ref{sec-evaluation} we conduct experiments with a dataset obtained 
from a real-world blog community and finally, conclude the paper in Section~\ref{sec-conclusions}.

\section{Relevant work}
\label{sec-related-work}

%% We mention various significant works, and explain why they are not directly relevant/useful to our problem
The recent explosion of Blogosphere has attracted a surge of research on issues related to Blogosphere modeling, mining, trust/reputation,
spam blog recognition, and many others~\cite{al08}; these issues though are not directly relevant to the present work. The specific problem
of identifying the influential bloggers in a blog site draws analogies from the problems of identifying influential blog sites and identifying 
authoritative Web pages (Web ranking). The identification of influential blog sites~\cite{gill1} and the related study of the spread of influence
among blog sites~\cite{gruhl1, gruhl2,java1,lkgfvg07} are orthogonal to the problem considered here, since we are interested in identifying 
influential bloggers in a single blog site, which might be or might not be an influential blog site. Similarly, the eigenvector-based methods for 
identifying authoritative Web pages~\cite{Langville-Google-book-2006}, like PageRank and HITS, ``{\it are not useful to our problem, since blog 
sites in Blogosphere are very sparsely linked}" \cite{kritikopoulos1}. Finally, it is obvious that the works which propose methodologies for 
discovering and analyzing blog communities~\cite{yuru1,yingzhou1} can not be exploited/tailored to our problem.

%% More details for the only competitor, emphasizing in its drawbacks
The only work directly relevant to our problem is that reported in~\cite{agarwal2}, which introduced the problem. To solve it, the authors
proposed an intuitive model for evaluating the blog posts. This model is based on four parameters: Recognition (proportional
to the incoming links), Activity Generation (proportional to the number of comments), Novelty (inversely proportional to the outgoing links) and
Eloquence (inversely proportional to the post's length). These parameters are used to generate an influence graph in which the influence
flows among the nodes. Each node of this graph represents a single blog post characterized by the four aforementioned properties. An influence
score is calculated for each post; the post with maximum influence score is used as the blogger's representative post. The influence score~$I(p)$
of a blog post~$p$ that is being referenced by~$\iota$ posts and cites~$\theta$ external posts, is determined by the following equation:
\begin{equation}
\label{agarwal-eq}
I(p)=w({\lambda})(w_{com}{\gamma}_{p}+w_{in}\sum_{m=1}^{|\iota|}I_p(m)-w_{out}\sum_{n=1}^{|\theta|}I_p(n))
\end{equation}
where~$w({\lambda})$ is a weight function depending on the length~$\lambda$ of a post and~$w_{com}$ denotes a weight that can be used to
regulate the contribution of the number of comments (${\gamma}_{p}$). Finally, $w_{in}$ and~$w_{out}$ are the weights that can be used to
adjust the contribution of incoming and outgoing influence respectively. The calculation of this influence score is recursive (positive
reinforcement from incoming links and negative reinforcement from outgoing links), similar to the PageRank definition. This score is the
${\iota}Index$ metric, which can be later used to identify the most influential bloggers. Isolating a single post to identify whether a
blogger is influential or not, is an oversimplistic approach, and so it would be if they have used gross metrics, like average, median and
so on. A blogger may have published only a handful of influential posts and numerous others of low quality, whereas other bloggers may have published several tens of influential blogs only, whose score though is lower than the score of the most influential
blog of the former blogger. Therefore, the productivity of bloggers is a significant issue that has been overlooked by this preliminary model.

Another drawback of this preliminary model is that its output depends highly on user defined weights. The value change of the above 
properties can lead to different rankings. Hence, its outcome is not objective, as tuning the appropriate weights the model identifies
influential bloggers with different characteristics. In other words this model cannot provide a satisfactory answer to the question
\textit{``who is the most influential blogger?''}; but it can give answers to questions of type~\textit{``who is the most influential
blogger according to the number of comments that his/her posts received?''}.

But most importantly, this model (and also the naive models which are based on the~$k$ most active bloggers), ignore one of the most important 
factors in Blogosphere: Time. As already known~\cite{al08}, the Blogosphere expands at very high rates, as new bloggers enter the 
communities and some others leave it. Hence, an effective model that identifies influential bloggers, should take into consideration the date 
that a post was submitted and the dates that the referencing posts were published, in order to be able to identify the {\it now-influential
bloggers}. Additionally, with such requirements it is mandatory to have fast methods (even on-line methods) for the discovery of the
influentials, which precludes the use of demanding and unstable recursive definitions, like that used by the influence-flow method proposed
in~\cite{agarwal2}.

\section{New metrics for evaluating\\the impact of blog posts}
\label{sec-scientometrics}

In this section we present new methods to assign influence scores to the blog posts of a blogger. These scores that will be used later
to identify the influentials. At first, we argue about what the desirable properties of these scores should be, and then we provide
the formulae for their calculation.

\subsection{Factors measuring a blogger's influence}
\label{subsec-influence-factors}

Beyond any doubt, the number of incoming links to a blog post is a strong evidence of its influence. Similarly, the number of comments made to
a post is another strong indication that this blog post has received significant attention by the community. The case of outlinks is more
subtle. In Web ranking algorithms like PageRank and HITS, the links are used only as a recognition of (or to convey)
authority. The influence-flow method of~\cite{agarwal2} assigns two semantics to a link: it is the means to convey authority, and also it the
means to reduce the novelty. This mechanism results in two significant problems: a) it misinterprets the intention of the link creators, and
b) it causes stability and convergence problems to the algorithm for the influence score calculation. It is characteristic that the authors admit
(\cite[page~215]{agarwal2}) that the presence of outlinks in novel posts is quite common and it is used ``to support the post's explanations".
Therefore, we argue that the outlinks are not relevant to the post's novelty, and all links should have a single semantic, that of implying
endorsement (influence).

The temporal dimension is of crucial importance for identifying the influentials. The time is related to the age of a blog post and also to
the age of the incoming links to that post. An influential is recognized as such if s/he has written influential posts recently or if its
posts have an impact recently. In the former case, the time involves the age of the post (e.g., in days since the current day) and in the
latter case, the time involves the age (e.g., in days since the current day) of the incoming links to the post.

There is another observation evident by the analysis presented in~\cite{agarwal2}:  a lot of the influential bloggers were also active
(i.e., productive) bloggers (see Table~1 and Tables 3--5 of~\cite{agarwal2}). Although, productivity and influence do not coincide, there is
a quite strong correlation among them. Therefore, productivity should somehow be taken into account when seeking for influential bloggers.

\subsection{The novel influence scores}
\label{subsec-novel-influence-score}

Based on the requirements described in the previous subsection, we develop formulae to estimate the influence of a blog post. We summarize some
useful notation in Table~\ref{table-notation}.

\begin{table}[!hbt]
\small
\begin{tabular}{|l|l|}\hline
{\bf Symbol}     & {\bf Meaning}\\\hline
$BP(j)$          & the set of blog posts of blogger $j$\\\hline
$bp_j(i)$        & $i$-th blog post of blogger $j$\\\hline
$C_j(i)$         & the set of comments to post $i$ of blogger $j$\\\hline
$R_j(i)$         & set of posts referring (have link to) the $i$-th post\\
                 & of blogger $j$\\\hline
$\Delta TP_j(i)$ & time interval (in days) between current time and\\
                 & the date that $j$-th blogger's post $i$ was submitted\\\hline
$\Delta TP(x)$   & time interval (in days) between current time and\\
                 & the date that post $x$ was submitted\\\hline
\end{tabular}
%\vspace*{-\baselineskip}
\caption{Notation.}
\label{table-notation}
\end{table}

As already mentioned, the map in Blogosphere changes rapidly, in a manner that a blogger who would currently considered as an influential, 
is not guaranteed to remain influential in the future. New bloggers enter the community and thousands of posts are submitted every day. In 
Section~\ref{sec-evaluation} it is demonstrated that a blogger may submit up to hundreds (or even thousands) of posts yearly. In this dynamic 
environment, the date that a blogger's post was submitted is crucial, since a blog post becomes ``old'' very quickly. An issue being discussed 
in a blog post at the present time and is now of major importance, may be totally outdated after two months. To account for this, we assign
a score $S^m_j(i)$ to the $i$-th post of the $j$-th blogger as follows:
\begin{equation}
S^m_j(i) = \gamma(|C(i)| + 1)(\Delta TP_j(i) + 1)^{-\delta} |R_j(i)|
\label{bcontemp-score}
\end{equation}
The parameter~$\gamma$ is not absolutely necessary, but it is used to grant to the quantities $S^m_j(i)$ a value large enough to be meaningful. 
Similarly, parameter~$\delta$ does not affect the relative score values in a crucial way, but it is used to allow for fast decaying of older
posts. Both parameters do not need complicated tuning, since they are not absolutely necessary; in our experiments, $\gamma$ and~$\delta$ are 
assigned values equal to~$4$ and~$1$, respectively. Since a post may receive no comments at all, we add one to the factor that counts the number 
of comments, to prevent null scores.

Using the definition of scores $S^m_j(i)$, we introduce a new metric \textit{MEIBI}\footnote{Metric for Evaluating and Identifying a 
Blogger's Influence.} for identifying influential bloggers. The definition of MEIBI follows:

\textit{Definition 1. }
%
%\begin{definition}
\label{def-meibi}
A blogger~$j$ has MEIBI index equal to~$m$, if~$m$ of his/her $BP(j)$ posts get a score $S^m_j(i) \geq m$ each, and the 
rest~$BP(j) - m$ posts get a score of $S^m_j(i) \leq m$.
%\end{definition}

This definition awards both influence and productivity of a blogger. Moreover, a blogger will be influential if s/he has posted several
influential posts recently.

But an old post may still be influential. How could we deduce this? Only if we examine the age of the incoming links to this post. If a post
is not cited anymore, it is an indication that it negotiates outdated topics or proposes outdated solutions. On the other, if an old post
continues to be linked to presently, then this is an indication that it contains influential material. Based on the ideas developed for the 
MEIBI metric, we work in an analogous fashion. Instead of assigning to a blogger's old posts smaller scores depending on their age, we can 
assign to each incoming link of a blogger's post a smaller weight depending on the link's age. This idea is quantified into the following
equation:
\begin{equation}
S^x_j(i) = \gamma(|C(i)| + 1)\sum_{\forall{x}\in{R_j(i)}}{(\Delta TP(x) + 1)^{-\delta}}
\label{btrend-score}
\end{equation}

Based on equation~\ref{btrend-score} the definition of the \textit{MEIBIX (MEIBI eXtended)} metric is formulated as follows:

\textit{Definition 2. }
%
%\begin{definition}
\label{def-meibix}
A blogger~$j$ has MEIBIX index equal to~$x$, if~$x$ of his/her $BP(j)$ posts get a score $S^x_j(i) \geq x$ each, and the 
rest~$BP(j) - x$ posts get a score of $S^x_j(i) \leq x$.
%\end{definition}

The introduction of the MEIBI and MEIBIX generates a straightforward policy for evaluating the influence of both blog posts and bloggers. 
No user-defined weights need to be set before these metrics provide results, whereas the most sound features of Blogosphere are considered. 
Moreover, the calculation of the metrics can be performed in an online fashion, since they do not involve complex computation and they do
not present stability problem like those encountered when using eigenvector-based influence scores. Note that the developed metrics are 
similar in spirit with the $h$-index and its variations (see~\cite{hirsch-wiki}) that recently became popular in the scientometrics 
litareture, but the challenges in Blogosphere are completely different: there are comments associated with each blog post, the time 
granularity is finer, the author of a post is a singe person, the resulting graph might contain cycles, and many more.

There is also the
possibility of taking into account the time that each comment was written, but such an extension does not contribute significantly to the
strength of the model, since the time-varying interest to the post is captured by the time-weighting scheme to the incoming links, and
moreover, it introduces the problem of having to handle two time scales, i.e., days for the links and the posts themselves, and hours or
minutes for the comments. In the sequel, we will evaluate the effectiveness of the proposed metrics to a real-world dataset, comparing it
with its only competitor~\cite{agarwal2}.

\section{Experimental evaluation}
\label{sec-evaluation}

The evaluation of the methods proposed here, but in general, of a lot others developed in the context of information retrieval, is tricky,
because there is no ground truth to compare against; things are more challenging in this case, since there is only alternative~\cite{agarwal2} to 
contrast with. Nevertheless, we firmly believe that our evaluation is useful and solid as long as the proposed methods reveal some latent 
facts that are not captured by the competitor and by some straightfoward methods, which result in different rankings for the final influential
bloggers. In the sequel of this section, we first describe the real data we collected for the experiments, and then present the actual
experiments and the obtained results.

\subsection{Data characteristics}
\label{subsec-collected-data}

Millions of blog sites exist. The Technorati\footnote{http://technorati.com} blog search engine claims to have indexed more than~115 million 
blogs. Since it is impossible to crawl the entire Blogosphere to obtain a complete dataset, it is essential to detect an active blog community 
that provides blogger identification, date and time of posting, number of comments and outlinks. The Unofficial Apple 
Weblog\footnote{http://www.tuaw.com} (TUAW) is a community that meets all these requirements; the same source of data was used also
in~\cite{agarwal2}. Although we use data from only one blog, the proposed methods can be appplied to every blog community having  characteristics similar to these of TUAW. We crawled\footnote{First week of December 2008.} TUAW and collected approximately~160
thousand pages, from which we extracted~17831 blog posts authored by~51 unique bloggers. This accounts for approximately~350 posts per
blogger on average. Moreover, the posts received totally~269449 comments (15 comments 
per post on average); only~1761 posts (ratio 10\%) were left uncommented. To obtain the incoming links to each
blog post, we used the Technorati API\footnote{http://technorati.com/developers/api/cosmos.html}. Apart from the number of the incoming links, 
we also retrieved the date that the referring post was submitted and its author's name. This information is necessary for the calculation of
the MEIBI and MEIBIX metrics. From the total~17831 blog posts, only~4586 of them had incoming links.
%(a ratio of 26\%) were cited by other blog posts. In details, 53575 incoming links derived from 6655 unique blog sites were collected.
Table~\ref{table-time-distrib} depicts the time distribution of both the blog posts and the incoming links.

\begin{table}[!hbt]
\begin{center}
\small
\begin{tabular}{|c|c|c|c|}\hline
{\bf Year} & {\bf Posts} & {\bf Posts with inlinks} & {\bf Inlinks} \\\hline
   2008    &    3676     &            3653          &     53204     \\\hline
   2007    &    4497     &             662          &       259     \\\hline
   2006    &    4354     &             186          &        18     \\\hline
   2005    &    4307     &              77          &         1     \\\hline
   2004    &     997     &               8          &         0     \\\hline
   Total   &   17831     &            4586          &     53575     \\\hline
\end{tabular}
\end{center}
%\vspace*{-1.25\baselineskip}
\caption{Time distribution of posts and inlinks.}
\label{table-time-distrib}
\end{table}

It is interesting to note, that 80\% of the total posts which have received at least one incoming link
(3653 posts out of the total~4586), were submitted within the year~2008. Consequently, either TUAW was not so popular before~2008 and the
bloggers were unaware of the information published there, or the posts submitted before~2008 were of medium or low quality, so that only
a few other bloggers referred to them. Hence, time-aware influence metrics which measure time difference in days, are indeed necessary
to differentiate between influential bloggers.

We investigate also the temporal distribution of the incoming links for a blog post measuring the intermediate time between the date a post
was submitted and the date it received each of the incoming links. The results are depicted in Table~\ref{table-citations-age}. Almost half
of the total inlinks were received (published) the same day that the post was submitted. Only a percentage of 2.3\% of all inlinks are dated one or more
years after the publication of the post. These results prove the necessity of time-aware metrics for the identification of the
influentials; since the posts are influential for a few days, {\it it is not particularly useful to identify influentials for the whole
lifetime of the blog site, but it is more substantial to identify the now-influential bloggers of the blog site}.

\begin{table}[h!]
\begin{center}
\small
\begin{tabular}{|l|c|c|}\hline
{\bf Age}                 & {\bf Inlinks} & {\bf Percentage}\\\hline
0 days                    & 26346         & 49,2\% \\\hline
1 day                     & 13470         & 25,1\% \\\hline
between 1 and 7 days      &  6653         & 12,4\% \\\hline
between 7 and 30 days     &  2406         & 4,5\% \\\hline
between 30 and 60 days    &   928         & 1,7\% \\\hline
between 60 and 365 days   &  2523         & 4,7\% \\\hline
over 365 days             &  1249         & 2,3\% \\\hline
Total                     & 53575         & 99,9\% \\\hline
\end{tabular}
\end{center}
%\vspace*{-1.25\baselineskip}
\caption{The age of the incoming links with respect to the publication date of the post they cite.}
\label{table-citations-age}
\end{table}

\subsection{Identifying the influential bloggers}
\label{subsec-experiment-influentials}

In this subsection we apply the proposed methods on the acquired dataset.
Apart from the proposed methods, we also examine a naive method which ranks the bloggers by using only their activity, i.e.,
number of published posts -- the activity index, one ranking method which is a straightforward adaptation of a method coming from the bibliometric
literature -- the h-index~\cite{hirsch-wiki} (we call these two methods as the plain methods), and a more sophisticated
method, proposed in~\cite{agarwal2}.

We divide the experimentation into three parts: in the first part, we compare the influential bloggers
indicated by the proposed methods, to the bloggers found by the plain methods. We use the entire dataset as a baseline experiment, examine whether temporal considerations are worthy examining; in the second part, we compare the influential bloggers indicated by the proposed
methods, with those found by the influence-flow method using the posts published in November~2008, to
prove that even for small time intervals the rankings are different; finally, we examine the temporal evolution of the influential bloggers
identified by the proposed methods during the year~2008, to examine whether the most influential bloggers lose their lead in influence and
strengthen even more the necessity for temporal considerations.

\subsubsection{The new methods vs. the plain ones}
\label{subsubsec-naive-methods}

Table~\ref{table-active} includes the ten most influential bloggers based solely on their activity (i.e., productivity) measured
by the number of posts they have published in TUAW. We also provide the dates that the first post (fourth
column) and the last post (fifth column) of each blogger was published. Although~\textit{S. McNulty} is ranked first,
he has not submitted any posts during the last~4.5 months. A similar observation of inactivity holds also for other top-10 influential
bloggers, like~\textit{D. Chartier} who is inactive in the last 3.5 months and~\textit{C.K. Sample, III}, who has no posts in the last 1.5 year. Recall, that both~\textit{S. McNulty} and~\textit{D. Chartier}, were ranked among the top-5 influential bloggers with the information-flow method (\cite[Table~1]{agarwal2}).

\begin{table}[h!]
\begin{center}
\small
\begin{tabular}{|c|l|c|c|c|}\hline
{} & {\bf Bloggers}       & {\bf $N$} & {\bf First}&{\bf Last}\\\hline
1  & S. McNulty        &    3037     &06/01/2005&31/07/2008\\\hline
2  & D. Caolo           &    2242     &07/06/2005&04/12/2008\\\hline
3  & D. Chartier       &    1835     &26/08/2005&30/08/2007\\\hline
4  & E. Sadun   		  &    1560     &09/11/2006&26/09/2008\\\hline
5  & C.K. Sample III      &    1057     &01/03/2005&05/06/2006\\\hline
6  & M. Lu               &    1043     &13/12/2006&04/12/2008\\\hline
7  & L. Duncan        &     954     &19/09/2004&23/01/2007\\\hline
8  & C. Bohon           &     793     &24/02/2004&04/12/2008\\\hline
9  & M. Rose         &     793     &29/11/2006&05/12/2008\\\hline
10 & M. Schramm	      &     648     &07/06/2007&04/12/2008\\\hline
%11 & Barb Dybwad          &     529     & 05/11/2004 & 05/03/2005 \\\hline
%12 & Sean Bonner          &     449     & 30/01/2004 & 25/08/2004 \\\hline
%13 & Robert Palmer        &     354     & 06/05/2008 & 04/12/2008 \\\hline
%14 & Victor Agreda, Jr.   &     314     & 01/02/2005 & 05/12/2008 \\\hline
%15 & Steven Sande         &     304     & 06/05/2008 & 04/12/2008 \\\hline
%16 & Damien Barrett       &     252     & 22/10/2005 & 01/08/2006 \\\hline
%17 & Brett Terpstra       &     226     & 28/12/2007 & 24/11/2008 \\\hline
%18 & Jan Kabili           &     186     & 30/08/2005 & 02/09/2006 \\\hline
%19 & Fabienne Serriere    &     148     & 22/10/2005 & 29/04/2006 \\\hline
%20 & Dan Lurie            &     143     & 10/06/2006 & 10/07/2007 \\\hline
\end{tabular}
\end{center}
%\vspace*{-1.25\baselineskip}
\caption{Bloggers ranking based on the number of posts submitted (active bloggers).}
\label{table-active}
\end{table}

Table~\ref{table-h-index} presents a ranking of the ten most influential bloggers when the h-index~\cite{hirsch-wiki} metric is used;
recall that this metric examines the number of posts of each blogger and the number of incoming links to each posts, awarding both productivity
and influence. The third column of Table~\ref{table-h-index} displays the value of the h-index metric for each blogger and the next two columns
show the total number of posts he/she has submitted in TUAW and how many of them have been cited by other posts respectively. Finally, the last
column illustrates the total number of incoming links that all the posts of a blogger have received.

\begin{table}[h!]
\begin{center}
\small
\begin{tabular}{|c|l|c|c|c|c|}\hline
{} & {\bf Bloggers} & {\bf h} & {\bf Posts} & {\bf Cited} & {\bf Inlinks} \\\hline
1 & E. Sadun & 31 & 1560 & 489 & 5759 \\\hline
2 & C. Bohon & 29 & 793 & 676 & 9439 \\\hline
3 & M. Schramm & 25 & 648 & 339 & 4322 \\\hline
4 & R. Palmer & 25 & 354 & 354 & 4809 \\\hline
5 & M. Rose & 24 & 793 & 364 & 4222 \\\hline
6 & D. Caolo & 23 & 2242 & 459 & 4907 \\\hline
7 & M. Lu & 23 & 1043 & 397 & 4282 \\\hline
8 & S. McNulty & 23 & 3037 & 334 & 3212 \\\hline
9 & B. Terpstra & 22 & 226 & 223 & 3013 \\\hline
10 & C. Warren & 22 & 133 & 112 & 1605 \\\hline
%11 & Steven Sande & 22 & 304 & 304 & 3775 \\\hline
%12 & Nik Fletcher & 18 & 128 & 72 & 1048 \\\hline
%13 & Chris Ullrich & 14 & 54 & 31 & 475 \\\hline
%14 & Victor Agreda, Jr. & 14 & 314 & 68 & 700 \\\hline
%15 & Joshua Ellis & 13 & 28 & 28 & 431 \\\hline
%16 & Giles Turnbull & 12 & 41 & 40 & 440 \\\hline
%17 & Jason Clarke & 9 & 14 & 12 & 144 \\\hline
%18 & Lisa Hoover & 9 & 30 & 16 & 196 \\\hline
%19 & TUAW Blogger & 8 & 11 & 11 & 113 \\\hline
%20 & David Chartier & 7 & 1835 & 113 & 279 \\\hline
\end{tabular}
\end{center}
%\vspace*{-1.25\baselineskip}
\caption{Bloggers ranking based on the h-index.}
\label{table-h-index}
\end{table}

Comparing Table~\ref{table-h-index} to Table~\ref{table-active}, some significant differences derive. These differences justify that productivity
and influence do not coincide. The most active blogger, \textit{S. McNulty} is ranked $8^{th}$ when the ranking is done in decreasing
h-index order. According to the h-index metric, the most influential blogger is \textit{E. Sadun} who has~31 articles that has at least~31
incoming links each. \textit{E. Sadun} is the fourth most active blogger in TUAW, though she has posted nothing in the last 2.5 months. Although she has been inactive recently, she is still the most influential according to the h-index metric. This proves that the
h-index can indicate the most influential blogger, but cannot identify bloggers who are
\textit{both} influential and active.

In the sequel, we apply the two proposed metrics MEIBI and MEIBIX in our dataset. The ranking of the bloggers according to the MEIBI
metric is displayed in Table~\ref{table-contemporary-index}.

\begin{table}[h!]
\begin{center}
\small
\begin{tabular}{|c|l|c|c|}\hline
{} & {\bf Bloggers}   & {\bf $m$} & {\bf $C_j$}\\\hline
1  & C. Bohon       & 49        & 14745 \\\hline
2  & R. Palmer    & 46        & 9916 \\\hline
3  & S. Sande     & 36        & 7246 \\\hline
4  & E. Sadun      & 34        & 32432 \\\hline
5  & M. Rose     & 30        & 13499 \\\hline
6  & M. Schramm     & 30        & 12838 \\\hline
7  & C. Warren & 28        & 4857 \\\hline
8  & D. Caolo       & 27        & 27985 \\\hline
9  & M. Lu           & 25        & 17966 \\\hline
10 & B. Terpstra   & 17        & 3770 \\\hline
%11 & Scott McNulty & 17 & 41813 \\\hline
%12 & Victor Agreda, Jr. & 14 & 23359 \\\hline
%13 & Chris Ullrich & 11 & 1824 \\\hline
%14 & Nik Fletcher & 10 & 3033 \\\hline
%15 & Giles Turnbull & 9 & 1382 \\\hline
%16 & TUAW Blogger & 8 & 2640 \\\hline
%17 & Joshua Ellis & 7 & 627 \\\hline
%18 & Jason Clarke & 6 & 347 \\\hline
%19 & Lisa Hoover & 6 & 822 \\\hline
%20 & Alberto Escarlate & 3 & 437 \\\hline
\end{tabular}
\end{center}
%\vspace*{-1.25\baselineskip}
\caption{Bloggers ranking based on the MEIBI index.}
\label{table-contemporary-index}
\end{table}

The data displayed in Table~\ref{table-contemporary-index} indicate that the blogger whose posts were the most influential recently, is
\textit{C. Bohon}. This is partially explained by the fact that~676 out of the total~793 posts, have received~9439 references; it is
the highest number of incoming links among the other bloggers. Furthermore, all posts have been commented~14745 times.

On the other hand, \textit{E. Sadun}, the most influential blogger according to the h-index metric, falls in the fourth position; considering
the fact that she has remained relatively inactive in the past 2.5 months, this is a satisfactory result. \textit{R. Palmer} and \textit{S. Sande}
occupy the second and third position respectively.
All top-three bloggers have submitted posts within December~2008. This is an indication that the MEIBI index not only identifies the most
influential bloggers, but also the most active. It is a metric that suits very well to our case, as Blogosphere changes rapidly and our
metric manages to keep track of these changes by handling the ages of the posts and the comments that they receive.

Table~\ref{table-trend-index} presents the most influential bloggers according to the MEIBIX index. One may detect several similarities
between Table~\ref{table-contemporary-index} and Table~\ref{table-trend-index}. The most active blogger of TUAW, \textit{S. McNulty},
is not among the top-10 influential bloggers when the ranking is performed according to either MEIBI or MEIBIX. This indicates that
although ~\textit{S. McNulty} is undoubtedly an active blogger, he has not submitted influential posts recently.
Table~\ref{table-h-index} though, reveals that the blogger in question, is the $8^{th}$ most influential when the ranking is determined
by the plain h-index metric.

\begin{table}[h!]
\begin{center}
\small
\begin{tabular}{|c|l|c|}\hline
{} & {\bf Bloggers} & {\bf $x$} \\\hline
1 & C. Bohon & 48 \\\hline
2 & R. Palmer & 47 \\\hline
3 & S. Sande & 37 \\\hline
4 & E. Sadun & 33 \\\hline
5 & C. Warren & 30 \\\hline
6 & M. Rose & 29 \\\hline
7 & M. Schramm & 27 \\\hline
8 & M. Lu & 26 \\\hline
9 & D. Caolo & 25 \\\hline
10 & B. Terpstra & 15 \\\hline
%11 & Scott McNulty & 15 \\\hline
%12 & Victor Agreda, Jr. & 13 \\\hline
%13 & Nik Fletcher & 10 \\\hline
%14 & Chris Ullrich & 9 \\\hline
%15 & Giles Turnbull & 9 \\\hline
%16 & TUAW Blogger & 8 \\\hline
%17 & Joshua Ellis & 7 \\\hline
%18 & Jason Clarke & 6 \\\hline
%19 & Lisa Hoover & 6 \\\hline
%20 & Alberto Escarlate & 3 \\\hline
\end{tabular}
\end{center}
%\vspace*{-1.25\baselineskip}
\caption{Bloggers ranking based on the MEIBIX index.}
\label{table-trend-index}
\end{table}

Finally, we computed the correlation of the rankings produced by h-index, 
MEIBI and MEIBIX by using the~\textit{Spearman's rho} metric. The results
(Table~\ref{table-rho-1}) indicate that MEIBI and MEIBIX produce similar
rankings, but both of them diverge from the h-index ordering significantly.

\begin{table}[h]
\begin{center}
\small
\begin{tabular}{|l|c|}\hline
{\bf Methods} & {\bf $\rho$} \\\hline
h-index -- MEIBI  & 0.478788 \\\hline
h-index -- MEIBIX & 0.321212 \\\hline
MEIBI -- MEIBIX   & 0.951515 \\\hline
\end{tabular}
\end{center}
%\vspace*{-1.25\baselineskip}
\caption{Corellation of rankings}
\label{table-rho-1}
\end{table}

\subsubsection{The new methods vs. the influence-flow method}
\label{subsubsec-top-10}

For the comparison of the proposed metrics against the basic competitor, i.e., influence-flow method~\cite{agarwal2}, we select a subset of the
real data in order to be fairer. It was obvious by the experimentation of the previous paragraphs, that the inactivity has a dramatic effect upon
the final ranking. The real question concerning the usefulness of the proposed methods is whether in a small period of time, say a month, these
methods would provide different rankings than those of the influence-flow method. Thus, we selected to work upon the blog posts of 
November~2008 only. For comparison purposes, we present in Table~\ref{tab-compare-metrics} the top-10 of active (most productive) bloggers during
November~2008 as this ranking is provided by the TUAW site itself.

\begin{table*}
\begin{minipage}[b]{0.33\linewidth}
\centering
\small
\begin{tabular}{|c|l|c|c|c|}\hline
{} & {\bf Bloggers}     & {\bf $N$} & {\bf Inlinks} & {\bf $C_j$} \\\hline
1  & C. Bohon	        & 47          & 508             & 556       \\\hline
2  & R. Palmer        & 42          & 339             & 491       \\\hline
3  & S. Sande         & 34          & 354             & 177       \\\hline
4  & M. Schramm       & 29          & 203             & 166       \\\hline
5  & D. Caolo         & 20          & 163             & 178       \\\hline
6  & M. Rose          & 19          & 138             & 154       \\\hline
7  & B. Terpstra      & 15          & 103             & 87        \\\hline
8  & C. Warren        & 8           & 80              & 331       \\\hline
9  & M. Lu            & 8           & 71              & 248       \\\hline
10 & V. Agreda        & 5           & 30              & 42        \\\hline
\end{tabular}
\end{minipage}
\begin{minipage}[b]{0.21\linewidth}
\centering
\small
\begin{tabular}{|c|l|}
\hline
{} & {\bf Blogger} \\\hline
1 & C. Bohon       \\\hline
2 & R. Palmer      \\\hline
3 & M. Lu          \\\hline
4 & C. Warren      \\\hline
5 & D. Caolo       \\\hline
6 & C. Ullrich     \\\hline
7 & S. Sande       \\\hline
8 & M. Rose        \\\hline
9 & V. Agreda     \\\hline
10 & Jason Clarke  \\\hline
\end{tabular}
\end{minipage}
\begin{minipage}[b]{0.22\linewidth}
\centering
\small
\begin{tabular}{|c|l|c|}
\hline
{} & {\bf Blogger} & {\bf $m$} \\\hline
1 & C. Bohon       & 26        \\\hline
2 & R. Palmer      & 20        \\\hline
3 & S. Sande       & 20        \\\hline
4 & D. Caolo       & 17        \\\hline
5 & M. Schramm     & 16        \\\hline
6 & M. Rose        & 13        \\\hline
7 & M. Lu          & 8         \\\hline
8 & B. Terpstra    & 7         \\\hline
9 & C. Warren      & 7         \\\hline
10 & V. Agreda     & 4         \\\hline
\end{tabular}
\end{minipage}
\begin{minipage}[b]{0.22\linewidth}
\centering
\small
\begin{tabular}{|c|l|c|}
\hline
{} & {\bf Blogger} & {\bf $x$} \\\hline
1  & C. Bohon      & 27        \\\hline
2 & S. Sande       & 20        \\\hline
3 & R. Palmer      & 19        \\\hline
4 & D. Caolo       & 18        \\\hline
5 & M. Schramm     & 16        \\\hline
6 & M. Rose        & 13        \\\hline
7 & M. Lu          & 8         \\\hline
8 & B. Terpstra    & 7         \\\hline
9 & C. Warren      & 7         \\\hline
10 & V. Agreda     & 4         \\\hline
\end{tabular}
\end{minipage}

\caption{Bloggers ranking according to: TUAW (left). Influence-flow model (center). MEIBI and MEIBIX (right).}
\label{tab-compare-metrics}
\end{table*}

In Table~\ref{tab-compare-metrics} we present the most influential bloggers for November~2008 as they are provided by the influence-flow
method and the MEIBI and MEIBIX metrics. Neither MEIBI nor MEIBIX generate rankings that agree with the
TUAW ranking of bloggers. TUAW concerns \textit{R. Palmer} as more influential than~\textit {S. Sande}. On the other hand, MEIBI
concerns~\textit{R. Palmer} and~\textit{S. Sande} to be equally influential. The former has authored more posts which received more
comments, whereas the latter's posts although fewer, have been referenced more times by other posts. The ranking produced by MEIBIX
positions~\textit{S. Sande} into the second place, higher than ~\textit{R. Palmer}. We could state that MEIBIX is more sensitive to the
number of incoming references than MEIBI.

Comparing the rankings produced by the proposed methods with the ranking according to the influence-flow model, we can state that this model
assigns to \textit{C. Bohon} the first position of the list. The model concerns \textit{R. Palmer} as the second most influential blogger
for the period of November of~2008 and agrees with TUAW. Despite \textit{S. Sande} has published more articles that received more incoming
links, \textit{M. Lu}'s posts have attracted more comments. Hence, we conclude that \textit{M. Lu} is primarily influential inside the
TUAW community, whereas \textit{S. Sande} has published influential posts that stimulated other bloggers to refer to them.

\textit{D. Caolo} has authored less posts than \textit{S. Sande}. Although his articles attracted both less comments and inlinks, the
influence-flow model assigns him a higher rank than \textit{S. Sande}. Obviously, the model's determination of influential bloggers, by taking into consideration only the best post and discarding all others, leads to erroneous rankings.

The~\textit{Spearman's rho} metric was used to compute the correlation of the rankings of Table~\ref{tab-compare-metrics}.
The results illustrated in Table~\ref{table-rho-2}, reveal that MEIBI and MEIBIX produce rankings that diverge significantly
from the one generated by the influence-flow model.

\begin{table}[h]
\begin{center}
\small
\begin{tabular}{|l|c|}\hline
{\bf Methods} & {\bf $\rho$} \\\hline
TUAW -- influence-flow model     & 0.284848 \\\hline
TUAW -- MEIBI     & 0.948485 \\\hline
TUAW -- MEIBIX    & 0.939394 \\\hline
influence-flow model -- MEIBI    & 0.418182 \\\hline
influence-flow model -- MEIBIX   & 0.357576 \\\hline
MEIBI -- MEIBIX   & 0.987879 \\\hline
\end{tabular}
\end{center}
\caption{Corellation of rankings}
\label{table-rho-2}
\end{table}

\subsubsection{Temporal evolution of the rankings produced by MEIBI and MEIBIX}
\label{subsubsec-temporal-evol-meibi}

\begin{figure}[!h]
\centering
\epsfig{file=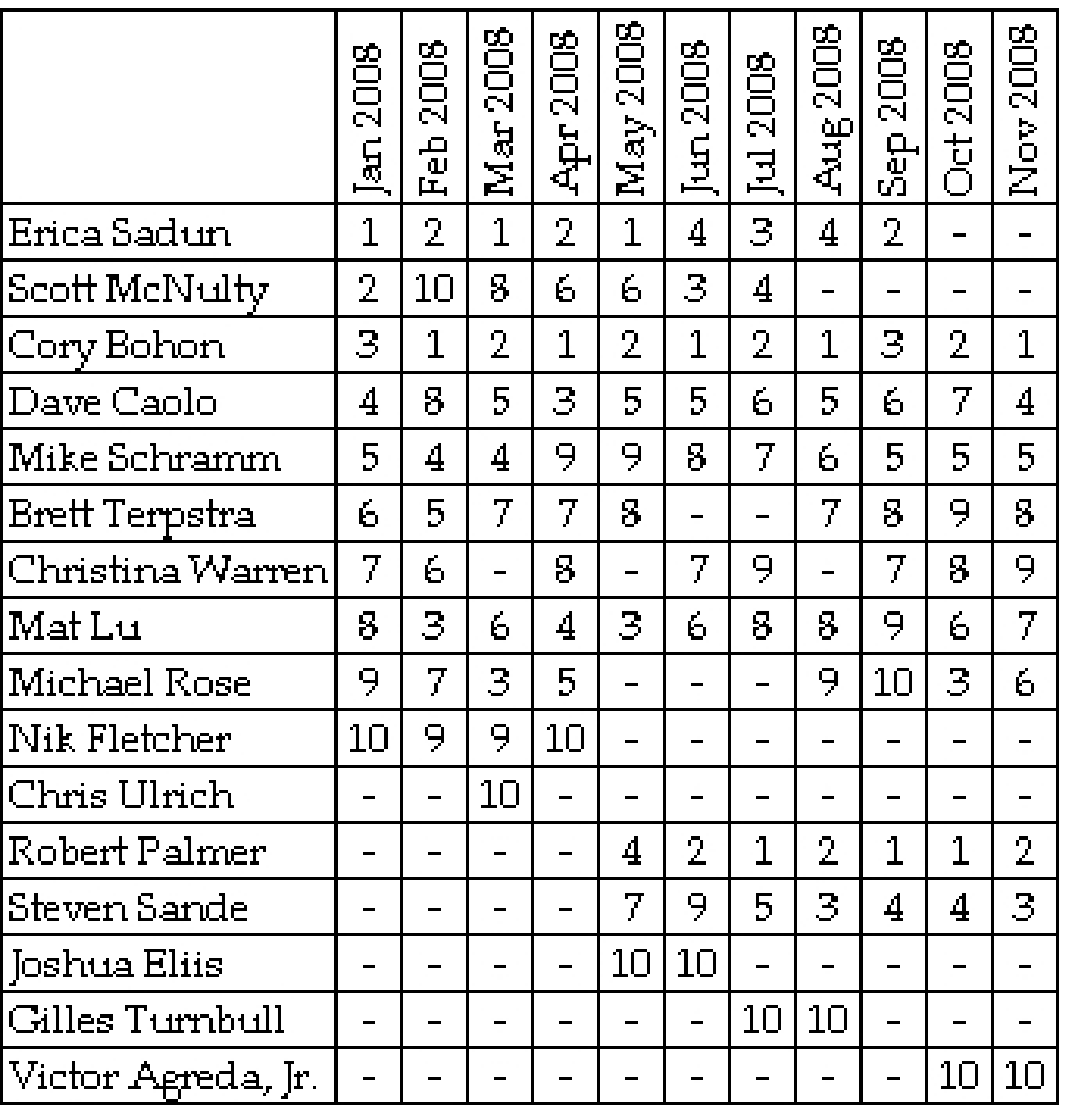, width=2.0in}
\caption{Influential bloggers' blogging behavior over 2008, according to MEIBI.}
\label{fig-meibi-11mon}
\end{figure}

\begin{figure}[!hbt]
\centering
\epsfig{file=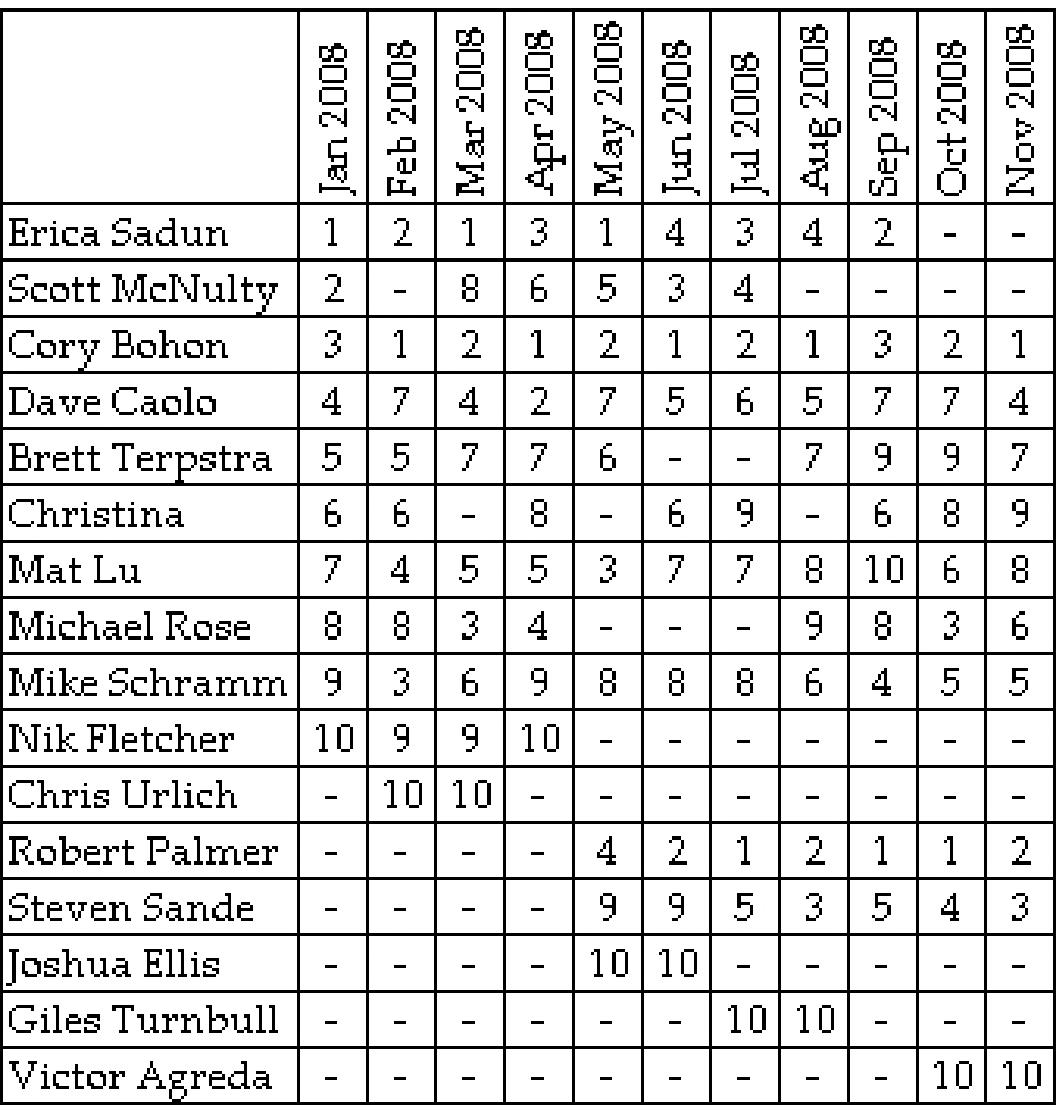, width=2.0in}
\caption{Influential bloggers' blogging behavior over 2008, according to MEIBIX.}
\label{fig-meibix-11mon}
\end{figure}

Finally, it is interesting to examine how the rankings generated by the proposed metrics vary over time. Figures~\ref{fig-meibi-11mon} and~\ref{fig-meibix-11mon}
depict the top-10 influence rankings of the bloggers in the past~11 months (from January~2008 to November~2008), when MEIBI and MEIBIX are applied
respectively. The columns in Figures~\ref{fig-meibi-11mon} and~\ref{fig-meibix-11mon} represent the progression of time, whereas the rows contain
the bloggers, ordered according to the time they were recognized as influential. Therefore, the $(i, j)$-th cell stores the rank of the $i^{th}$
blogger in the $j^{th}$ time window. The dash symbol signifies that the particular blogger was not among the top-10 of that period.

MEIBI and MEIBIX produce similar rankings; MEIBIX is more affected by the number of incoming links, whereas MEIBI assigns better scores to
the posts that attracted more comments.

Studying the blogger rankings fluctuation over time, composes a valuable tool for distinguishing bloggers that have been influential for a very
long or very short time. The former can be considered as more influential, as compared to the latter which are proved more trustworthy. Certainly,
many other categories of bloggers can be derived from the retrospection of their activity through time and many potential applications can be
developed using these categories.

\section{Conclusions}
\label{sec-conclusions}

The Blogosphere has recently become one of the most favored services on the Web. Many users maintain a blog and write posts to express
their opinion, experience and knowledge about a product, an event, and several others comment upon these opinions. This ``participatory
journalism" of blogs has such an impact upon the masses that Keller and Berry~\cite{keller1} argued that through blogging ``one American in tens tells
the other nine how to vote, where to eat and what to buy''. Therefore, a significant issue is how to identify such influential bloggers,
because commercial companies can turn the influentials to become their ``unofficial spokesmen", innovative business opportunities related
to commercial transactions and traveling can be developed capitalizing upon the influentials, and so on.

This article investigated the problem of identifying influential bloggers in a blog site and proposed two new methods that provide
rankings of the influentials. The main motivation for the introduction of these methods is that the closely relevant, competing methods
have not taken into account temporal aspects of the problem, which we argue are the most important ones when dealing with spaces like
the Blogosphere, which is highly volatile and doubles in size every six months.

The first proposed metric, termed MEIBI, takes into consideration the number of the blog post's inlinks and its comments, along with the
publication date of the post. The second metric, MEIBIX, is used to score a blog post according to the number and age of the blog post's
inlinks and its comments. The metrics can be computed very fast because they do not involve complex recursive definitions of influence,
and in addition they do not use tunable parameters which are difficult to set. Therefore, they can be used in an online fashion for the
identification of the {\it now-influential bloggers}.

These methods were evaluated against the state-of-the-art influential blogger identification method, namely that reported in~\cite{agarwal2},
utilizing data collected from a real-world community blog site. The obtained results attested that the new methods are able to better identify
significant temporal patterns in the blogging behaviour, and reveal some latent facts about the blogging activity.

\end{document}